\documentclass[conference]{IEEEtran}
\usepackage{fancyhdr}

\IEEEoverridecommandlockouts
\usepackage{cite}

\usepackage{amsmath,amssymb,amsfonts}
\usepackage{algorithmic}
\usepackage{float}
\usepackage{graphicx}
\usepackage[hidelinks]{hyperref}
\usepackage{listings}
\usepackage{makecell}
\usepackage{orcidlink}
\usepackage{subcaption}
\usepackage{textcomp}
\usepackage{url}
\usepackage{verbatim}
\usepackage{xcolor}

\definecolor{lbcolor}{rgb}{0.9,0.9,0.9}
\definecolor{Darkgreen}{rgb}{0,0.4,0}
\def\BibTeX{{\rm B\kern-.05em{\sc i\kern-.025em b}\kern-.08em
    T\kern-.1667em\lower.7ex\hbox{E}\kern-.125emX}}
\begin{document}

\title{Porting and Benchmarking Chapel on Emerging RISC-V Hardware: An HPC Viability Study
}

\author{
\IEEEauthorblockN{
Ian Henriksen\IEEEauthorrefmark{1},
Chris Taylor\IEEEauthorrefmark{2},
Patrick Diehl\IEEEauthorrefmark{3},
Jade Abraham\IEEEauthorrefmark{4},
Palmer Cox\IEEEauthorrefmark{5}, \\
Bradford L. Chamberlain\IEEEauthorrefmark{4}, and
Stephen L. Olivier\IEEEauthorrefmark{1}
}

\IEEEauthorblockA{\IEEEauthorrefmark{1}
Sandia National Laboratories, Albuquerque, NM, USA\\
\{idhenri, slolivi\}@sandia.gov}

\IEEEauthorblockA{\IEEEauthorrefmark{2}
Tenstorrent (currently), Santa Clara, CA, USA\\
Tactical Computing Labs (formerly), Lindsay, TX, USA\\
ctaylor@tenstorrent.com}

\IEEEauthorblockA{\IEEEauthorrefmark{3}
Los Alamos National Laboratory, Los Alamos, NM, USA\\
diehlpk@lanl.gov}

\IEEEauthorblockA{\IEEEauthorrefmark{4}
Hewlett Packard Enterprise, Seattle, WA, USA\\
\{jade.abraham, bradford.chamberlain\}@hpe.com}

\IEEEauthorblockA{\IEEEauthorrefmark{5}
FiscalNote, Washington DC, USA\\
p@lmercox.com}

\IEEEauthorblockA{
\small
ORCID:
Ian Henriksen: 0000-0002-2053-7265;
Chris Taylor: 0000-0001-7119-818X;\\
Patrick Diehl: 0000-0003-3922-8419;
Jade Abraham: 0009-0002-9880-579X;\\
Palmer Cox: 0009-0000-6697-7242;
Stephen L. Olivier: 0000-0001-6247-8980;\\
Bradford L. Chamberlain: 0000-0002-6065-2049
}
}

\bstctlcite{BSTcontrol}

\maketitle


\begin{abstract}
The Chapel programming language recently added support for the RISC-V architecture.
Here we discuss what changes were needed for Chapel to work on RISC-V as well as lessons learned from the porting process.
We use some of Chapel's extensive benchmark suite to gain further insight into the suitability of the RISC-V architecture for future HPC use.
We compare performance on SiFive P550 and Unmatched boards and a Sophon SG2042 with a variety of other recent HPC CPU platforms.
Various portability and performance anomalies arose on different architectures and are discussed.
While the RISC-V machines represent smaller preliminary offerings and not HPC-class hardware, these results still provide hope that RISC-V hardware can become viable for HPC in the near future.
\end{abstract}

\begin{IEEEkeywords}
Chapel, RISC-V, High Performance Computing (HPC), SiFive P550, Sophon SG2042, User-level Threads, Qthreads
\end{IEEEkeywords}

\setlength{\textfloatsep}{10pt plus 1.0pt minus 2.0pt}
\setlength{\intextsep}{10pt plus 1.0pt minus 2.0pt}
\setlength{\dbltextfloatsep}{10pt plus 1.0pt minus 2.0pt}
\setlength{\dblfloatsep}{10pt plus 1.0pt minus 2.0pt}

\section{Introduction}
The RISC-V ISA \cite{riscv-spec} has seen increasing adoption and more widespread software support in recent years.
HPC-class processors have been announced by NextSilicon~\cite{nextsilicon2026arbel}, Tenstorrent~\cite{tenstorrent-ascalon}, SiFive~\cite{sifive-p870}, and others.
RISC-V is already seeing preliminary use in accelerator cards targeting HPC and AI workloads\cite{nextsilicon-maverick}\cite{inspiresemi-thunderbird}\cite{tenstorrent-wormhole}.
Nvidia is already shipping RISC-V cores in large quantities\cite{nvidia-riscv} as internal components of other products.

In the interest of gauging the readiness of RISC-V for use in HPC, we ported Chapel \cite{chamberlain2015} (and the Qthreads threading runtime \cite{wheeler2008} that it uses) to RISC-V.
This provided insight into both the difficulty of porting a large HPC code to work on RISC-V and the capability of available RISC-V hardware.
To do this performance evaluation we ran a series of well-maintained Chapel benchmarks on the Sophgo Sophon SG2042 \cite{sg2042-specs}, HiFive P550 \cite{p550-specs}, and HiFive Unmatched, as well as a series of more widely known and well-studied ARM and x86 processors.

Porting Chapel to RISC-V demonstrated that adding support for a new ISA is only a moderate challenge for a well-maintained codebase that uses modern interfaces like C11 atomics \cite{c23-standard} to limit the amount of platform-specific assembly.
Greater availability of binary packages for Chapel's dependencies would have further helped ease this process; however, the recent RVA23 profile \cite{rva23} promises to make this more tractable going forward.
The changes to the Chapel and Qthreads build systems were minimal.
The primary change needed was adding a RISC-V assembly context swap implementation to Qthreads.
While adding preliminary support for RISC-V was still nontrivial, the effort involved was outweighed by the constant and ongoing load of maintaining good performance on newly released processors that use established ISAs.
Examples are provided for this discussion.
This effort highlights the need for further work on performance portability, especially as RISC-V enables a greater variety of hardware in the HPC space.

While the benchmark results did still favor established ISAs, substantial differences like cost and process node likely explain much of the disparity.
Remaining performance anomalies likely stem from known and/or resolved issues.
This gives hope that future chips that directly target the HPC market and server-class workloads can attain competitive performance.

Key contributions and insights from this study include:
\begin{itemize}
\item Porting Chapel and its HPC runtime to RISC-V.
\item Many specific lessons learned from the port.
\item Performance evaluation of the CLBG benchmarks and the ChOp optimization framework~\cite{carneiro2019} on RISC-V.
\end{itemize}

\subsection{Chapel}
\label{section:intro:chapel}
Chapel is an open-source parallel programming language designed to improve the productivity, portability, and scalability of high-performance computing applications~\cite{callahan2004cascade,chamberlain2015}.
It provides high-level abstractions for task and data parallelism, distributed-memory execution, and locality-aware programming while retaining explicit control over data placement and computation~\cite{deitz2006chapel}.
Chapel uses a global-view programming model in which parallelism can be expressed using a unified set of language constructs across shared-memory systems, distributed-memory clusters, and accelerators.
The compiler and runtime manage many low-level aspects of execution, including task scheduling, synchronization, communication, and distributed data management.
These characteristics make Chapel a useful platform for studying the challenges associated with porting a modern parallel language, compiler, and runtime system to an emerging architecture such as RISC-V.

\subsection{Qthreads}
\label{section:intro:qthreads}
The Qthreads library\cite{wheeler2008} provides lightweight threading with flexible full-empty bit synchronization.
The API allows decomposition of programs into many user-level threads (ULTs), or tasks, that are scheduled onto a smaller number of long-running pthreads (typically one per core or hardware thread).
Fast context swapping and the limited state of ULTs, compared to pthreads, makes overhead costs manageable.
Full-empty bits, in the past demonstrated in experimental parallel architectures~\cite{smith1981hep}, are here implemented in software to allow for event-driven execution.
Qthreads can be used directly through its C API, but it is perhaps most suited to use as a tasking or threading layer for higher-level systems like Chapel (where it is currently the default threading backend).

\subsection{Organization}
Section~\ref{section:porting} discusses what changes were required to get Chapel and Qthreads working on RISC-V and insights learned during that process.
Section~\ref{section:machines} describes the experimental setup for the benchmark runs.
Section~\ref{section:clbg} discusses the performance of the Chapel submissions for the Computer Language Benchmarks Game.
Section~\ref{section:chop} discusses the performance of ChOp\cite{carneiro2019} (a larger Chapel code) on RISC-V.
Section~\ref{section:smt-numa} discusses preliminary observations around how SMT and NUMA affect the different benchmarks run on the baseline x86 and ARM systems.
Section~\ref{section:related} discusses other related work.
Section~\ref{section:conclusion} concludes.

\section{Porting Chapel to RISC-V}
\label{section:porting}
Porting Chapel and Qthreads to work on RISC-V proved to be straightforward.
Not all possible build configurations are supported; however, some of the optimized configurations are available.
Currently, there is almost no support for RISC-V in the software packages for interacting with GPUs or distributed communication layers, so Chapel's support for RISC-V does not currently include distributed or GPU-based execution.
On a practical level, the upcoming availability of binary packages targeting the RVA23 profile~\cite{rva23} (and hardware that supports that profile) promises to further ease the process of standing up large software stacks on the RISC-V ISA.
There were, of course, a few changes to the Chapel and Qthreads build systems; however, these ended up being extremely small.
The main nontrivial code change was adding a RISC-V assembly implementation of fast context swaps to Qthreads.
There were also minor changes elsewhere to make things like cache-line size estimation and pausing during spinlocks work as expected.
The effort required for porting Chapel and Qthreads to RISC-V demonstrates that adding support for RISC-V is tractable for software that minimizes its use of assembly through careful organization and portability interfaces like C11 atomics.
Adding this support also allowed us to contrast the effort involved in adding initial support for RISC-V with the constant ongoing effort in maintaining performance on new high-performance chips that use more established architectures.
This comparison highlights both the cost and need for further work in performance portability.

\subsection{Build Configurations}
\label{section:porting:build-configurations}
Chapel now has preliminary support for RISC-V, but the choice of build configuration is important.
Some build configurations will not work due to optional third-party library dependencies that have not been ported.
There are also build configurations that would have worked with less effort, but they are for debugging and are not expected to perform well.

Chapel requires a parallel memory allocator for performance and supports multiple backends.
Jemalloc~\cite{jemalloc} has historically been used, but the most recent versions of Jemalloc that support RISC-V are not yet supported in Chapel due to changes in how the memory allocator must interact with underlying distributed communication layers.
These issues do not affect the shared-memory setting, but we still recommend using a different allocator on RISC-V for now.
Mimalloc\footnote{\url{https://github.com/microsoft/mimalloc}}, the allocator used for the experiments here, already supports RISC-V and is also expected to perform well.

Chapel can be built to run without Qthreads, but it is optimized around the low-cost threading and synchronization Qthreads provides.
The fallback work queues perform poorly.

For the experiments here, we also did not use Chapel's optional C-compiler backend.
It supports using a C compiler as a backend instead of LLVM, but this configuration is not the default.
It is an older compatibility layer with reasonable but less-tested performance.

\subsection{Runtime Support}
\label{section:porting:runtime-support}
Major Linux distributions have supported RISC-V for several years. 
Ubuntu has provided RISC-V images since Ubuntu 20.04 LTS in 2020, while experimental Fedora support dates back to Fedora 27 in 2018. 
Similarly, GCC and LLVM introduced RISC-V support in 2017 and 2018, respectively.
In our experience, compiling Chapel for RISC-V was straightforward.
However, the compilation times for both Chapel and LLVM were considerably longer than on server-class x86 and Arm systems.

Despite the availability of operating-system and compiler support, support for high-performance communication layers remains limited.
InfiniBand is supported by the Linux kernel, but currently available RISC-V systems often lack full-size PCIe slots, provide limited PCIe lane bandwidth, or have incomplete support for firmware, interrupts, IOMMUs, and DMA.
HPE Cray Slingshot could, in principle, be used with RISC-V, but its host software stack would need to be ported and validated for the architecture.
Consequently, until HPC-grade RISC-V platforms become widely available, most RISC-V systems are connected using conventional Ethernet, with MPI used for distributed-memory execution~\cite{diehl2024preparing,10.1145/3624062.3624230}.
For our experiments, we explicitly built Chapel in a shared-memory-only configuration.

GPU support on RISC-V is similarly immature.
NVIDIA announced in 2025 that it plans to support RISC-V host processors in the CUDA platform, but no public availability timeline was provided\footnote{\url{https://riscv.org/blog/nvidia-to-bring-cuda-platform-support-to-the-risc-v/}}.
Experimental support for AMD GPUs and ROCm has been demonstrated\footnote{\url{https://github.com/ROCm/ROCm/issues/5629}}, but is not official yet.
Intel Arc GPUs have been used in experimental RISC-V configurations, but RISC-V is not currently an officially supported host architecture for Intel oneAPI.
Because of this, we did not evaluate Chapel's GPU benchmarks.
This limitation reduced the number of benchmarks and real-world applications that could be included in our study.

\subsection{Binary Package Availability}
\label{section:porting:binary-packages}
In practice, one significant barrier to setting up Chapel on RISC-V systems was the lack of availability of binary packages.
While Chapel bundles copies of many of its third-party dependencies, Chapel depends on LLVM and various build tools.
Obtaining a correctly built copy of a recent version of LLVM when the underlying platform does not provide any kind of binary can be time-consuming.
As of this writing, Ubuntu 26.04\footnote{\url{https://canonical.com/blog/canonical-releases-ubuntu-26-04-lts-resolute-raccoon}} provides binary builds of various recent versions of LLVM; however, this is a recent development.
Support for RISC-V in the latest versions of Ubuntu is restricted to chips that implement the RVA23 Profile~\cite{rva23}; however, chips with that architecture are rapidly becoming available.

\subsection{Build System Modifications}
\label{section:porting:build-system}
Chapel includes a bespoke build system that connects third-party codes that use GNU autotools or CMake.
The only change needed was one line, correctly recognizing 64-bit Linux based on identifying \lstinline{riscv64} as the machine type.

The only change needed in the Qthreads build system was to include RISC-V in the list of architectures where fast context swap implementations are known to be available.
This prevents the build system from falling back to the slower system-provided context swaps from \lstinline{ucontext.h}.

\subsection{Context Swap Port}
\label{section:porting:context-swap}
The primary nontrivial code change to enable Qthreads and Chapel on RISC-V was implementing a fast context swap in RISC-V assembly for Qthreads.
Context swaps are used by operating systems to multiplex OS threads over the available cores.
They suspend the execution of the current thread and then jump to the stack of a new thread and continue executing wherever the new thread had left off.
User-level threading systems like Qthreads allow users to create threads that behave like OS threads but are generally more lightweight, maintaining significantly less state.
The user-level threading capability extends existing System V infrastructure provided in the Linux kernel.
User-level threads have cheaper context swaps and scheduling constructs due to fundamentally different underlying assumptions about security, signal management, address space, etc.
The Qthreads threading runtime handles multiplexing the work of large numbers of lightweight threads over the available hardware resources.

While Qthreads does offer the option to fall back to the system-provided default context swaps commonly available in \lstinline{ucontext.h}, this option is expected to perform poorly and should not be used for anything other than debugging.
In particular, the system-provided context swap is expected to handle things like signal masks and floating-point contexts even though those are semantically not required for swaps between qthreads.
Swapping the signal mask can only be done via syscall on Linux, making it more expensive on current hardware than any register shuffling done as a part of the context switch itself.
Because of this, Qthreads works better when using faster context swaps that do not change anything about the signal mask.

In order to provide a fast RISC-V context swap for use in Qthreads, it was necessary to:
\begin{itemize}
    \item Define a struct type to hold the necessary saved registers.
    \item Implement the RISC-V assembly to swap the current context to/from that struct (following the calling convention of \lstinline{swapcontext}).
    \item Implement a RISC-V version of \lstinline{makecontext} to initialize a context object.
    \item Connect these implementations so that they are used for the RISC-V architecture.
\end{itemize}


\subsection{CPUID}
\label{section:porting:cpuid}
Qthreads includes best-effort logic to guess the cache-line size based on the CPU ID.
For RISC-V, we get the cache-line size using \lstinline{sysconf} since glibc\footnote{\url{https://ftp.gnu.org/gnu/glibc/}} provides an option for doing so.
We fall back to 128 bytes as a conservative estimate if \lstinline{sysconf} does not provide an interface for inferring the L1 cache-line size (e.g., when using musl libc\footnote{\url{https://musl.libc.org/}}).

\subsection{Pause}
\label{section:porting:pause}
Qthreads also needed an equivalent for the pause instruction.
It semantically does nothing from the perspective of the C and C++ abstract machine and is not strictly required for correctness, but this instruction is used inside various spinlock-like constructs for performance reasons.
We provided the appropriate RISC-V equivalent for a pause in the appropriate Qthreads macro.

This requires the underlying processor to support the zihintpause extension \cite{riscv-spec}.
This extension is present on all RISC-V hardware available to us for testing and is included in the RVA23 profile \cite{rva23}, so we opted to require it.

\subsection{Atomics as a Portability Layer}
\label{section:porting:atomics}
A major factor in enabling support for RISC-V in Qthreads (and Chapel) was the recent removal of a large amount of platform-specific assembly code.
These changes were made independently before the port to RISC-V.
They were intended to improve the maintainability of Qthreads, but also simplified the work needed for porting Qthreads to new architectures.

C11 (and C++11) atomics (see \cite{c23-standard} for the latest revision) were designed as a portability layer to abstract away the architectural specifics around atomic accesses and memory fences.
They have been overwhelmingly successful in that regard, but they are still a low-level construct for building parallel software.
For example, the performance issues encountered in the Thread Ring benchmark (see Section~\ref{section:clbg:synchronization}) underscore that the performance may still vary dramatically.
The implications of specific atomic access and synchronization patterns are highly platform-specific.
Atomics are a portability layer, but they are not a performance portability layer.

\subsection{Performance Portability Status}
\label{section:porting:performance-portability}
A key takeaway we gained from the work porting Chapel and Qthreads to RISC-V has revolved around the relative effort of porting to a new architecture vs. the constant effort of maintaining performance across the architectures we currently support.
Now that Qthreads primarily uses C11 atomics as a portability interface, the effort required for adding a new architecture is modest.
It is still impossible to get around providing things like a fast context swap implementation and a \lstinline{pause} equivalent instruction; however, the changes beyond that are minor.

On the other hand, there is always a constant stream of performance bugs surfacing as updated hardware and software are released.
Even with established architectures like x86 and ARM, the effort required to maintain good performance as new chips are released is substantial.
For example, in the past few months, the Chapel team found and reported three separate upstream performance regressions in LLVM 22.
The Qthreads developers are currently investigating the anomalous performance on the Grace and Granite Rapids processors (see Section~\ref{section:clbg:synchronization}) and the highly variable startup time (see Section~\ref{section:clbg:no-op}).
The observed performance of the benchmarks shown here varies little across different LLVM versions, as can be seen in the supplementary data, but this is not automatic.
It comes because of continued ongoing efforts toward performance maintenance by the Chapel team.

Support for the RISC-V ISA itself was reasonable to add, and the SG2042 and P550 studied here appear to perform as expected, given their relative costs and known issues.
On the other hand, the extensibility and configurability of the RISC-V ISA opens the door for a much greater variety of hardware from a greater variety of vendors.
This highlights the urgent and expanding need for work around performance portability.
The performance bugs discussed here require a significant investment of time to fix.
They highlight the categorical difference between nominal support for an ISA and performance that is confirmed to work well for a specific machine.

\section{Machine Configurations}
\label{section:machines}
\begin{table*}[ht]
    \centering
    \caption{Machines used for benchmarking. TDP is CPU only. Cost numbers are estimates only, and it is noted whether they are for an SoC, whole machine, or CPUs only. For machines that have two CPUs, 2x the power draw and current CPU MSRP are reported. Since Nvidia only sells whole systems for the Grace-Grace CPU, the reported price is for a cheap whole system with a publicly listed price at the time of writing.}
    \label{table:hardware-stats}
    \addtolength{\tabcolsep}{-0.4em}
    \begin{tabular}{|l|lllllllll|}\hline
    Hardware & ISA & CPU & Cores & Clock & Memory & Process Node & Year & MSRP & TDP \\\hline
    MILK-V Pioneer & RISC-V & Sophon SG2042 \cite{sg2042-specs} & 64 & 2.4GHz & 128GB & TSMC 12nm \cite{strack2026} & 2023 & \$2200 (SoC) & 120W  \\
    HiFive Premier P550 & RISC-V & Eswin EIC7700X \cite{p550-specs} & 4 & 1.4GHz & 16GB & TSMC 12nm & 2024 & \$399 (SoC) & 8W \cite{p550-datasheet} \\
    HiFive Unmatched Rev B & RISC-V & SiFive Freedom U740 & 5 & 1.2GHz & 16GB & TSMC 28nm & 2024 & \$299 (SoC) & 5W \cite{unmatched-cluster} \\\hline
    Nvidia Grace-Grace & ARM & 2x Grace \cite{grace-specs} & 144 & 3.1GHz & 512GB & TSMC 4N \cite{grace-blog} & 2023 & \$38400 (whole) & 500W \\\hline
    Intel Granite Rapids & x86 & 2$\times$ Xeon 6980P \cite{xeon6980p-specs} & 256 & 2.0GHz & 1500GB & Intel 3 & 2024 & \$27910 (CPUs) & 1000W \\
    AMD Zen 3 & x86 & 2$\times$ EPYC 7763 \cite{epyc7763-specs} & 128 & 2.45GHz & 1000GB & TSMC 7nm \cite{zen-core-architecture} & 2021 & \$15780 (CPUs) & 560W \\
    Intel Ice Lake & x86 & 2$\times$ Xeon 8352Y \cite{xeon8352y-specs} & 64 & 2.2GHz & 256GB & Intel 10nm & 2021 & \$7990 (CPUs) & 410W \\\hline
    \end{tabular}
    \addtolength{\tabcolsep}{+0.4em}
\end{table*}

\begin{table}[tb]
    \centering
     \caption{Software used for benchmarking}
    \label{table:software-setup}
    \begin{tabular}{|l|lll|}\hline
    Hardware & OS & Kernel & LLVM  \\\hline
    MILK-V Pioneer & Fedora 38 & 6.1.55 & 21.1.8 \\
    HiFive Premier P550 & Ubuntu 24.04 & 6.12.33 & 22.1.0 \\
    HiFive Unmatched Rev B & NixOS 26.05 & 7.1.3 & 21.1.8 \\
    Nvidia Grace-Grace & Rocky Linux 9.7 & 5.14.0 & 21.1.8 \\
    Intel Granite Rapids & Rocky Linux 8.10 & 4.18.0 & 21.1.8 \\
    AMD Zen 3 & Rocky Linux 8.10 & 4.18 & 21.1.8 \\
    Intel Ice Lake & Rocky Linux 9.6 & 5.14 & 21.1.8 \\
    \hline
    \end{tabular}
\end{table}

To get a sense for how the SG2042 and P550 perform relative to more established architectures, we tested a bevy of other recent server-class x86 and ARM processors with the various Chapel benchmarks.
The machines used for testing are shown in Table~\ref{table:hardware-stats} with data drawn from the cited announcements and spec sheets and additional missing data gathered from elsewhere \footnote{\url{https://chipsandcheese.com/p/inside-sifives-p550-microarchitecture}} \footnote{\url{https://www.theregister.com/offbeat/2020/10/29/sifive-inches-closer-to-offering-a-true-risc-v-pc-latest-five-core-dev-board-includes-pcie-ssd-interfaces/454977}} \footnote{\url{https://hothardware.com/reviews/sifive-hifive-unmatched-rev-b-review}}.
In the HPC space, server-class machines are of particular interest, so those are the machines used as a basis for comparison.
There are currently no comparable server-class RISC-V machines publicly available; however, the MILK-V Pioneer (SG2042) does have 64 cores.
The SG2042 is no longer available in the US due to trade restrictions; however, upcoming offerings from NextSilicon and Tenstorrent will be.
We also included the HiFive P550 machine as a point of reference that uses a somewhat newer core architecture.
The HiFive Unmatched board was included as a point of reference relative to older RISC-V hardware.

Among the x86 and ARM machines tested, the Grace and Granite Rapids machines were included in this comparison to show the relative performance of state-of-the-art x86 and ARM machines.
The Zen 3 and Ice Lake machines are somewhat older server-class systems that provide a better point of comparison for the RISC-V machines.
It should be noted that, despite being released two years earlier than the MILK-V, even the Ice Lake and Zen 3 system CPUs themselves cost dramatically more than the whole MILK-V Pioneer system and draw substantially more power.
Similarly, the MILK-V and P550 machines, despite being more recent releases, were made on inferior process nodes relative to the older Ice Lake and Zen 3 server-class processors.

Table~\ref{table:hardware-stats} includes some publicly available cost data for the CPUs in these systems.
While real-world costs depend heavily on organizational procurement details and the publicly available costs in Table~\ref{table:hardware-stats} vary depending on what type of cost data is publicly available, the differences are so dramatic that there are still meaningful comparisons to be made.
For example, the cost of just the two CPUs in the recent top-tier Granite Rapids system used for this comparison is more than an order of magnitude larger than the whole-system cost for the MILK-V Pioneer system which is, itself, the most expensive RISC-V system tested here.
Similarly, although the real-world power draw for a specific workload varies relative to the manufacturer's reported TDP, the Granite Rapids machine has a TDP of nearly an order of magnitude more than the SG2042.
The differences between server-class and consumer-class machines show up in dramatic ways in both cost and power use.

Performance results throughout the rest of the paper should be interpreted with the cost and power use in mind.
For HPC applications we are primarily interested in server-class systems.
For the comparisons here we are using the best publicly available RISC-V hardware we currently have access to; however, the differences in cost and power draw mean that the RISC-V performance is expected to be substantially worse in all cases.
On the other hand, the data provided here shows what the current state-of-the-art RISC-V hardware is capable of, and provides hope that many of the observed performance gaps can be closed with future server-class RISC-V chips.

Table~\ref{table:software-setup} lists the operating system and kernel versions.
LLVM 21 was compiled for use with Chapel on both the SG2042 and the P550.
Both LLVM and Chapel itself were cross-compiled for the Unmatched system.

The SG2042 does support a customized older variant (0.7.1) of the RVV spec; however, it is only available for use with the old custom vendor-provided GCC compiler\footnote{\url{https://github.com/XUANTIE-RV/xuantie-gnu-toolchain}}.
The custom GCC is version 7, which was too old to be suitable for the experiments here.

As noted in Section~\ref{section:porting:runtime-support}, no commercial GPU runtime is supported on RISC-V, and support for distributed communication is still limited.
Thus, these benchmarks are exclusively single-node and CPU-focused.
In Chapel terms, these benchmarks are all run within a single locale with no GPUs.

Performance results, unless otherwise noted, are reported as an average of 10 runs.
Log scales are used to allow displaying the results for the Unmatched system alongside the results for the Grace and Granite Rapids systems meaningfully in the same figures.
For brevity in the tables, ``Unmatched" is shortened to ``Unm", ``Granite Rapids" is shortened to ``GR", and ``Ice Lake" is shortened to ``IL".

\section{CLBG Benchmarks}
\label{section:clbg}
Chapel has put substantial effort into optimizing its performance on the Computer Language Benchmarks Game (CLBG) benchmark suite\footnote{\url{https://benchmarksgame-team.pages.debian.net/benchmarksgame/}}.
This is a benchmarking suite intended for testing the performance of various programming languages in a series of simple tasks.
Here, rather than testing the performance of different programming languages, we are using the optimized CLBG benchmarks included in Chapel to estimate the suitability of different computing architectures for HPC workloads, as exemplified by Chapel itself.

The CLBG benchmarks are an extensive and well-curated set of Chapel benchmarks that do not require GPUs or distributed execution.
They include startup time, so the associated costs are discussed in Section~\ref{section:clbg:no-op} and then the startup costs are subtracted out of the averages for the remaining sections.
Though some benchmarks like the Thread Ring and Chameneos benchmarks are no longer included in the CLBG benchmark suite, we opted to include them since they provide additional insight into synchronization costs on the architectures studied here.
Several of the CLBG benchmarks have multiple versions that have been submitted by the Chapel team.
These different versions are denoted with different numbers.

\subsection{Chapel Startup Time}
\label{section:clbg:no-op}
\begin{table}[h!]
\vspace{-5pt}
\caption{Chapel startup times (seconds)}
\label{table:noop}
\vspace{-5pt}
\begin{center}
\begin{tabular}{|c|c c c c c c c|}
\hline
Name & SG2042 & P550 & Unm & Grace & Zen 3 & GR & IL \\
\hline
No-Op & 0.14 & 0.02 & 0.04 & 0.08 & 1.36 & 0.67 & 0.10 \\
\hline
\end{tabular}
\end{center}
\vspace{-5pt}
\end{table}

While investigating the performance of the CLBG benchmarks across architectures, it became apparent that shorter benchmarks always ran better on the smaller machines.
This is because the CLBG benchmarks measure end-to-end program time and not just the amount of time spent on the benchmark's computation.
Larger machines have a higher startup time cost for Chapel.
To measure this, we included a set of benchmark runs of a no-op Chapel benchmark that measures the cost of Chapel runtime setup.
Investigation into why the startup time is so much higher for specific architectures is ongoing.
To simplify the discussion throughout the remainder of Section~\ref{section:clbg}, we have subtracted off the startup time from the remaining benchmark runs.

\subsection{Digits of Pi}
\label{section:clbg/pidigits}

\begin{figure}[h]
  \vspace{-12pt}
  \centering
  \includegraphics[width=\linewidth]{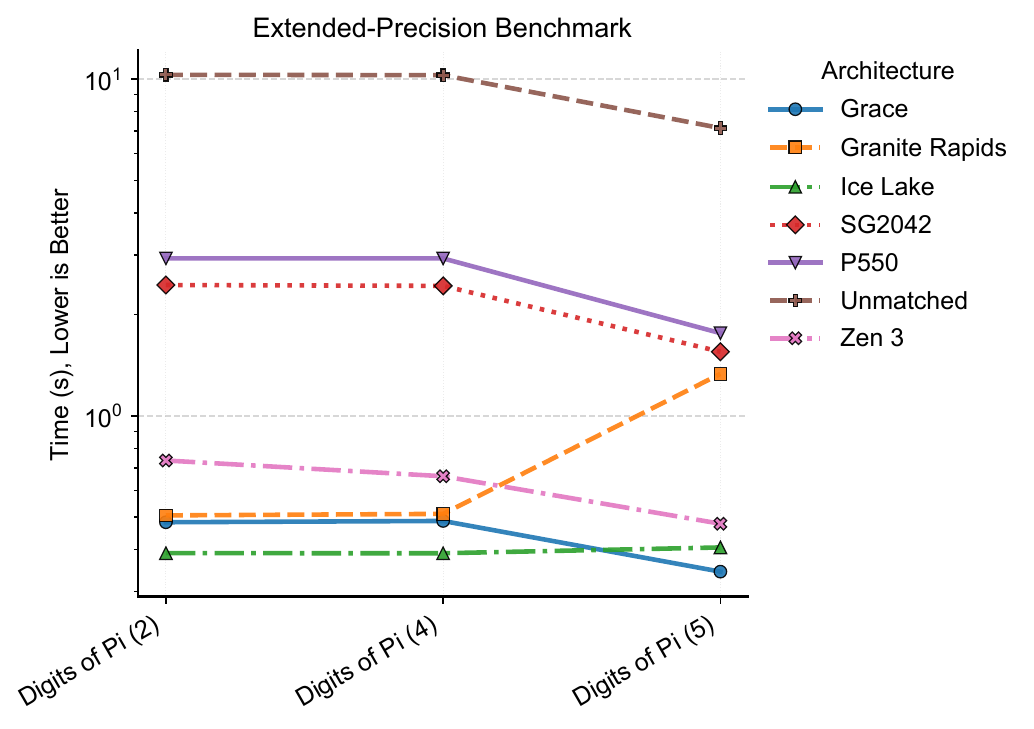}
  \caption{Digits of Pi Benchmark Performance}
  \label{figure:pidigits}
\end{figure}

``Digits of Pi" is a benchmark for computing the first $10000$ digits of $\pi$.
All the Chapel versions rely heavily on the GMP library\footnote{\url{https://gmplib.org}} for extended precision arithmetic.
Versions 2 and 4 execute sequentially, so they measure the performance of a single core, while version 5 uses 4-way pipeline parallelism.

The performance of the various platforms on this benchmark initially appears to indicate that the per-core performance with RISC-V cores is still substantially lower than state-of-the-art ARM and x86 cores.
The disparity may be larger in this case due to a known issue with the RISC-V ISA lacking a single instruction to perform a combined add-with-carry operation.
This operation is critical for implementing extended precision arithmetic and requires three separate instructions instead of one on RISC-V.
That said, extended precision arithmetic is a specialized usage pattern present only in extremely specific workloads.
Granite Rapids also performed poorly on the parallel version of this benchmark.

\subsection{Binary Trees (Allocation Intensive)}
\label{section:clbg/bintrees}

\begin{figure}[ht]
  \centering
  \includegraphics[width=\linewidth]{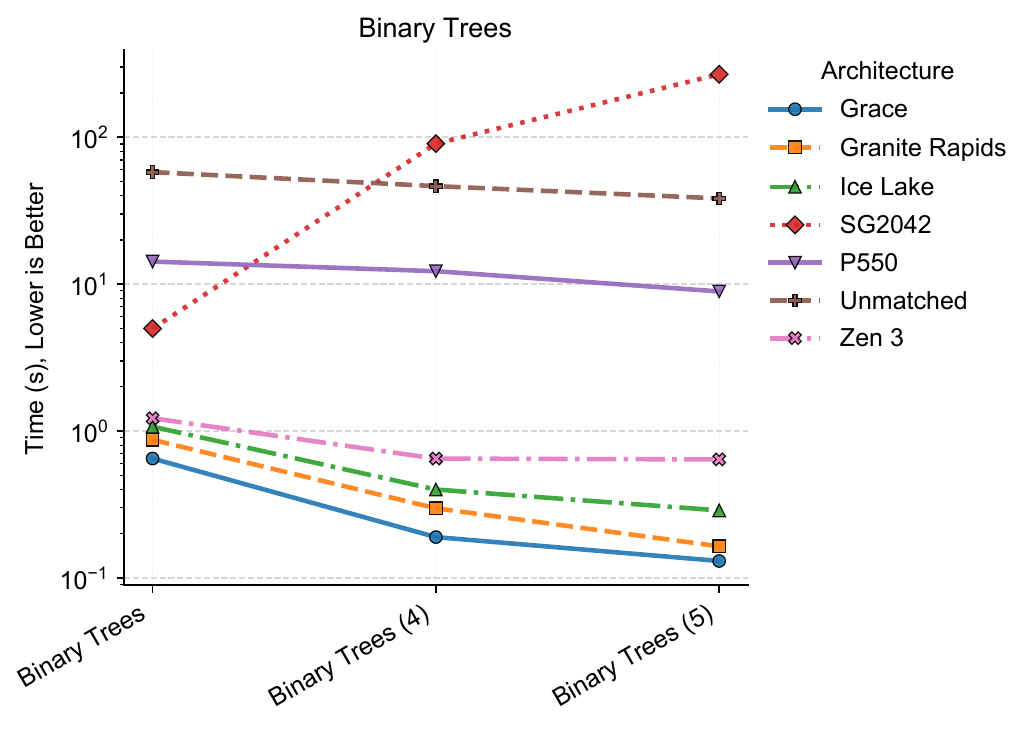}
  \caption{Binary Trees Benchmark Performance}
  \label{figure:binarytrees}
\end{figure}

The binary trees benchmark is a simplified version of an older garbage collection benchmark\footnote{\url{https://hboehm.info/gc/gc_bench/}}.
It involves building a series of perfect binary trees with constraints on how allocations can be performed.
It is especially demanding of the CPU's memory subsystem.

The results here show a dramatic slowdown for the SG2042 relative to the other platforms (even the P550 and Unmatched boards) on some versions of this benchmark.
This is unsurprising since the memory subsystem in that processor has known performance issues \cite{brown2024} \cite{brown2025}.
This issue was essentially resolved with the refreshed SG2044 chip (see \cite{brown2025}).
This issue has less effect on version 1 of this benchmark.
It is also not present for the other RISC-V chips.
In that case, the disparities are unsurprising given the dramatic differences in price, process node, and power consumption.

\subsection{IO-Intensive Benchmarks}
\label{section:clbg:io-intensive}

\begin{figure}[h]
  \vspace{-12pt}
  \centering
  \includegraphics[width=\linewidth]{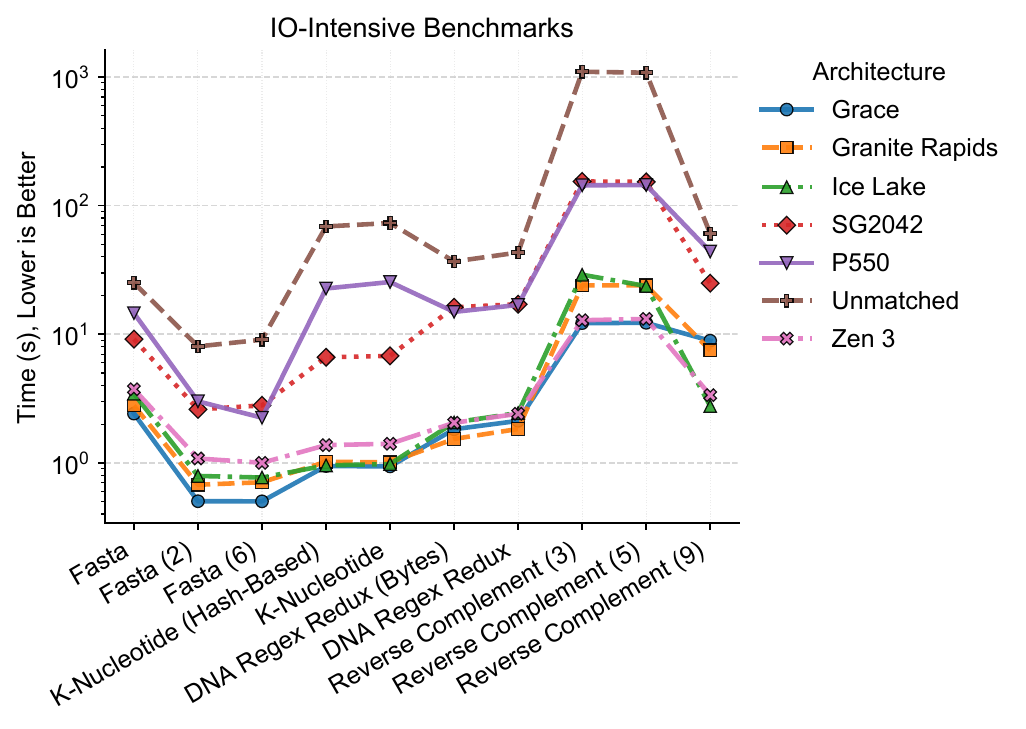}
  \caption{IO-Intensive Benchmark Performance}
  \label{figure:io}
\end{figure}


Several CLBG benchmarks involve nontrivial IO operations.
In particular, the Fasta, K-nucleotide, DNA Regex Redux, and Reverse Complement benchmarks all involve various simple manipulations on a large string of DNA data.
In spite of the nontrivial IO, the differences follow what would be expected given the disparities in cost and power use.

\subsection{Floating-Point-Intensive Benchmarks}
\label{section:clbg:fp-intensive}

\begin{figure}[ht]
  \centering
  \includegraphics[width=\linewidth]{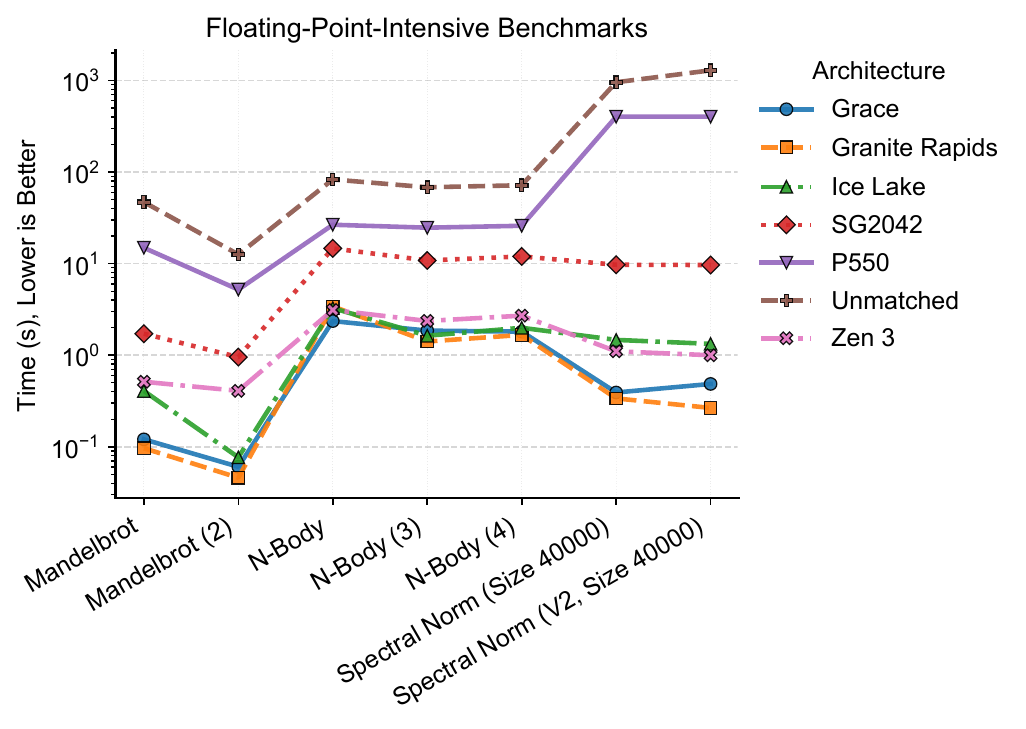}
  \caption{Floating-Point-Intensive Benchmark Performance}
  \label{figure:fp}
\end{figure}


The Mandelbrot benchmark computes a $16000\times16000$ image of a Mandelbrot set using a naive parallel algorithm then writes the result to file.
While it includes some IO, in the experience of the Chapel developers, it is very dependent on vectorized floating-point performance.
The performance gap between the RISC-V machines and the state-of-the-art Grace and Granite Rapids machines is pronounced in this case.
This is likely due to the lack of vectorization as none of the RISC-V machines tested support RVV 1.0.
The ARM and x86 machines all have access to relevant vector extensions.

The Spectral Norm benchmark computes a numerical estimate of the spectral norm of a large implicitly-stored matrix through repeated matrix-vector multiply operations.
Only the vector is ever fully materialized.
The matrix itself is dense, so a large number of floating-point operations are necessary.
The existing spectral norm benchmark works on a $5500\times5500$ matrix, but the running time for a problem that small is dwarfed by the program startup time.
To compensate for this, we introduced an additional version of the benchmark with a larger test-problem size of $40000\times40000$ to better compare between architectures.
This larger benchmark is now included in Chapel.
For the larger matrix size, the Grace and Granite Rapids machines dramatically outperform the other machines.
As was the case with the Mandelbrot benchmark, this disparity is likely due to the use of dedicated vector instructions.
Given that RVV is seeing increased adoption, there is hope that these sorts of disparities in floating-point performance can be resolved in the near future.

The N-Body simulation benchmark does a naive simulation of a small number of bodies for 50,000,000 time steps.
It still shows better performance on the server-class chips, indicating that there is room for improvement in the serial floating-point performance on these RISC-V machines, but the difference is less dramatic than with the other floating-point-related benchmarks.

\subsection{Synchronization Benchmarks}
\label{section:clbg:synchronization}

\begin{figure}[ht]
  \vspace{-12pt}
  \centering
  \includegraphics[width=\linewidth]{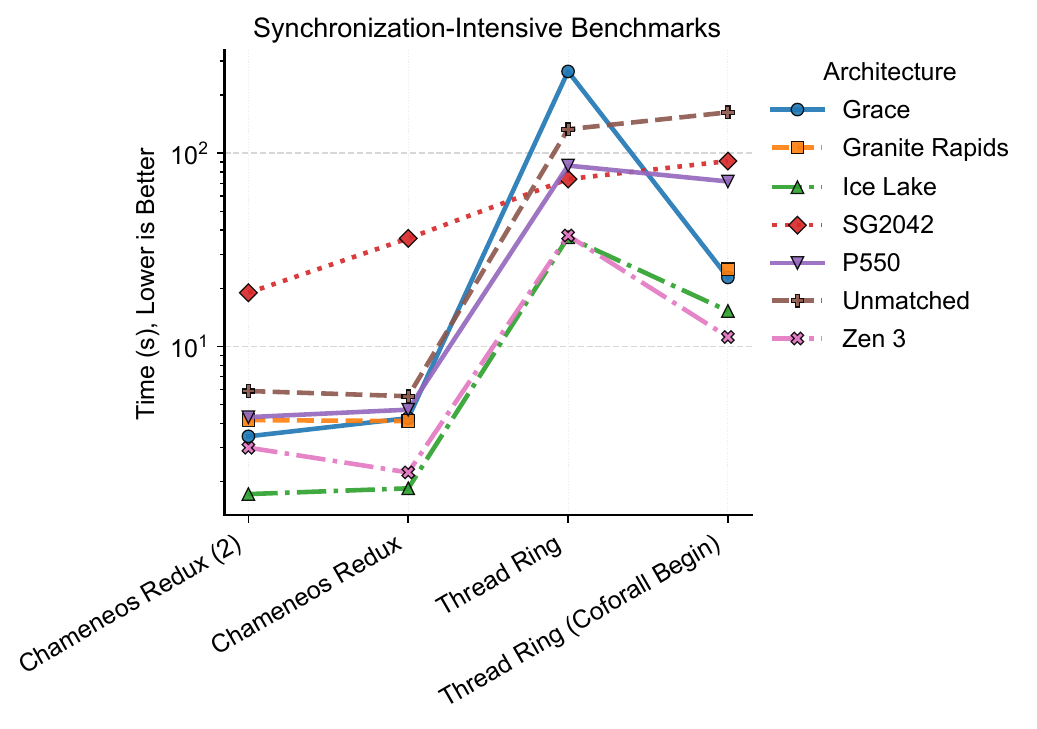}
  \caption{Synchronization Benchmark Performance 
  }
  \label{figure:synchronization}
\end{figure}

The Chameneos and Thread Ring benchmarks are former benchmarks from the CLBG.
They measure the speed of certain types of multithreaded synchronization patterns.
These benchmarks are inherently parallel in how they operate, but they do not have large embarrassingly parallel sections of work to hide synchronization overheads.
They are designed to measure the synchronization overheads themselves.

The Chameneos Redux benchmark\cite{chameneos} simulates a series of meetings between creatures where each waits for others to arrive and then updates its own state (color) based on the others that arrive.
$6000000$ meetings are simulated between a small population of creatures.

In the Thread Ring benchmark, 503 lightweight threads are created and then a token is passed between threads in a loop for a specified number of steps ($50000000$).
It measures the ability of the underlying runtime and architecture to suspend and then rapidly resume execution across many threads/cores.

In practice, these benchmarks favor the older server architectures and the smaller/simpler architectures.
This can be seen in Figure~\ref{figure:synchronization}.
The Zen 3 and Ice Lake machines consistently performed best.
The P550 showed performance close to the Grace and Granite Rapids machines for the Chameneos benchmark, while the SG2042 performed substantially worse than even the Unmatched board.
The standard Thread Ring benchmark shows dramatic slowdowns on both the Grace and Granite Rapids architectures.
Further inspection on the Granite Rapids machine showed that the running times varied wildly and frequently exceeded timeout limits.
We have excluded that data point entirely from the plot because of this.
Among the results remaining, Grace performed by far the worst, taking longer than even the Unmatched machine.

In the interest of gaining more information about the Thread Ring performance anomaly we ran Chapel's archived versions of the Thread Ring benchmark.
All but one of the older versions of the Thread Ring benchmarks showed similar anomalous behavior.
The thread-ring-coforall-begin benchmark did not, so we have included it here as an additional point of reference.
That version divides work up in a way that results in more lightweight threads and is less likely to exhibit contention on empty work queues.
Again, the performance favored older server-class machines for this benchmark, but the performance on the Grace and Granite Rapids machines was more reasonable.

Work to address the Thread Ring performance bug is ongoing.
The inconsistent performance here highlights the need for additional study of what kinds of atomic communication patterns and parallel idioms work well on emerging architectures, especially as core counts continue to increase and interconnects increase in complexity.


\subsection{Fannkuch Redux}
\label{section:clbg:fannkuch}
\begin{table}[h!]
\vspace{-5pt}
\caption{Fannkuch Redux performance (seconds)}
\label{table:fannkuch}
\vspace{-5pt}
\begin{center}
\addtolength{\tabcolsep}{-0.1em}
\begin{tabular}{|c|c c c c c c c|}
\hline
Name & SG2042 & P550 & Unm & Grace & Zen 3 & GR & IL \\
\hline
Fannkuch & 0.91 & 19.02 & 26.49 & 0.13 & 0.57 & 0.13 & 0.55 \\
\hline
\end{tabular}
\addtolength{\tabcolsep}{0.1em}
\end{center}
\vspace{-5pt}
\end{table}

The Fannkuch Redux benchmark \cite{fannkuch} involves repeatedly permuting small ranges of integers in an array.
It iterates over all possible permutations on sets of a given size (in this case $12$), and is embarrassingly parallel.
This benchmark shows good performance on the Grace and Granite Rapids architectures.
The SG2042 is less than 2x slower than the Intel Ice Lake and Zen 3 machines, showing some room for improvement, but also performing well relative to its cost.
The P550 and Unmatched boards perform poorly.
This makes sense given their lower core counts.

\subsection{Takeaways}
\label{section:clbg:takeaways}
The cutting-edge Grace and Granite Rapids systems generally performed best in the CLBG benchmarks except in the case of the synchronization benchmarks.
There, performance favored the older server-class machines, and to a lesser extent, the simpler RISC-V boards.
While the RISC-V machines we had available for these comparisons do not achieve competitive performance with state-of-the-art x86 and ARM machines, this is expected due to their dramatically lower cost and power use.
Specific improvements like the availability of hardware with support for the RVV extensions as well as the announcements of new RISC-V chips targeting the server-class market give hope that the performance gap can be closed.
In many ways this ongoing progress echoes the emergence of HPC-class ARM chips in that the initial hardware focused on the low-cost, low-power market segment and then spread to server-class and HPC workloads.

\section{ChOp}
\label{section:chop}
\begin{table}[t]
\caption{ChOp optimization library performance (seconds)}
\label{table:chop}
\begin{center}
\begin{tabular}{|c|c c c c c c c|}
\hline
Name & SG2042 & P550 & Unm & Grace & Zen 3 & GR & IL \\
\hline
ChOp & 49.28 & 998.27 & 1588.04 & 8.40 & 16.70 & 7.19 & 49.23 \\
\hline
\end{tabular}
\end{center}
\vspace{-5pt}
\end{table}

ChOp \cite{carneiro2019} is a branch-and-bound optimization engine written in Chapel.
It supports GPU execution \cite{carneiro2020}; however, since existing GPU runtimes do not yet work on RISC-V, performance data for the CPU-only version is presented here.
It is included in this benchmark to highlight that larger, nontrivial Chapel codes can also work on RISC-V systems.
To run this benchmark we used the ``queens\_singlelocale\_cpu'' version with size $19$ and initial depth $5$.

ChOp does best on the Grace and Granite Rapids machines; however, the SG2042 roughly matches the performance of the Ice Lake machine, performing excellently for its price.
The P550 and Unmatched machines underperform in this setting, in part because they have so many fewer cores.

\section{SMT and NUMA Effects}
\label{section:smt-numa}
\begin{figure*}[ht]
  \centering
  \begin{subfigure}[b]{0.48\textwidth}
    \includegraphics[width=\linewidth]{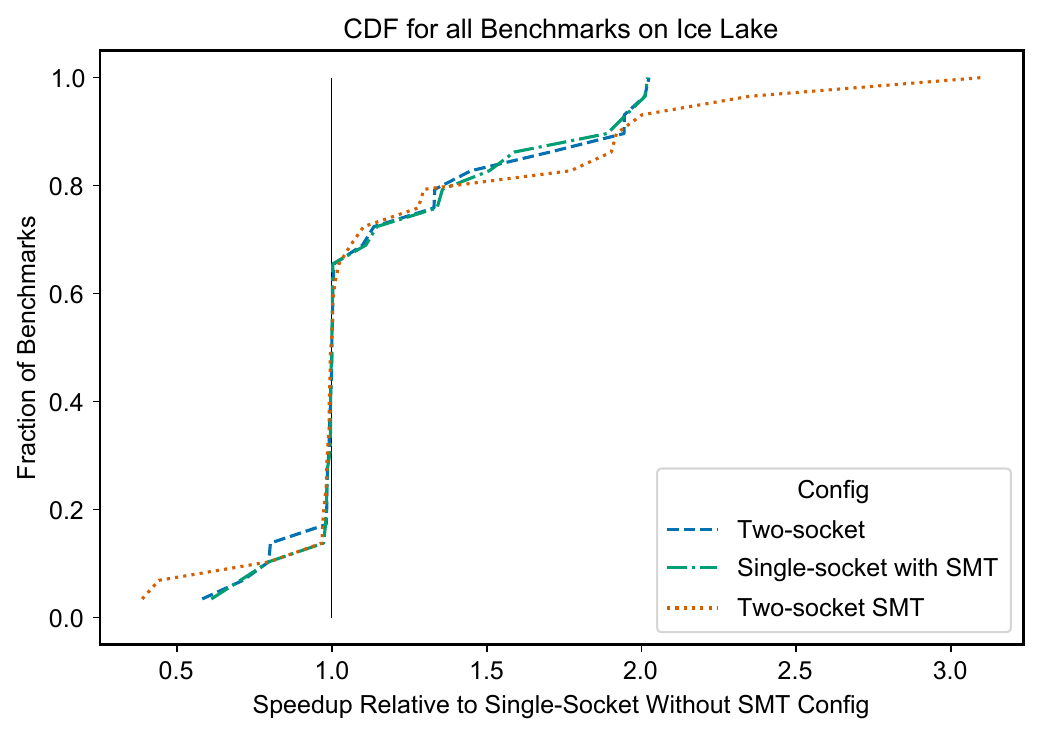}
  \end{subfigure}
  \begin{subfigure}[b]{0.48\textwidth}
    \includegraphics[width=\linewidth]{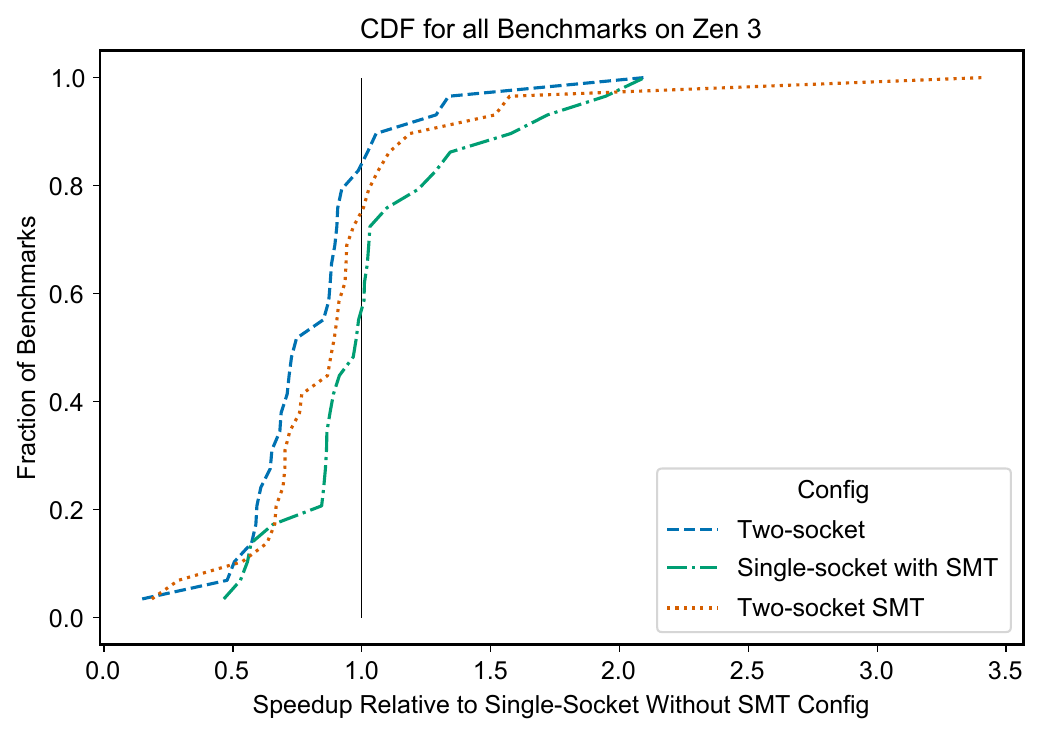}
  \end{subfigure}
  \begin{subfigure}[b]{0.48\textwidth}
    \includegraphics[width=\linewidth]{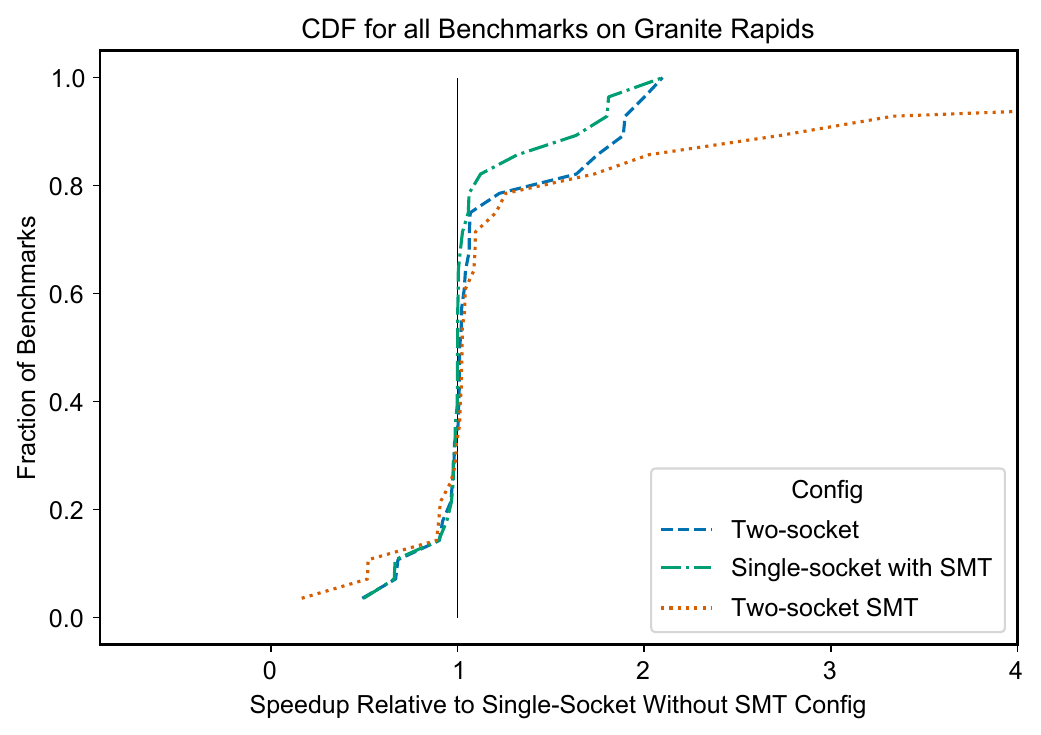}
  \end{subfigure}
  \begin{subfigure}[b]{0.48\textwidth}
    \includegraphics[width=\linewidth]{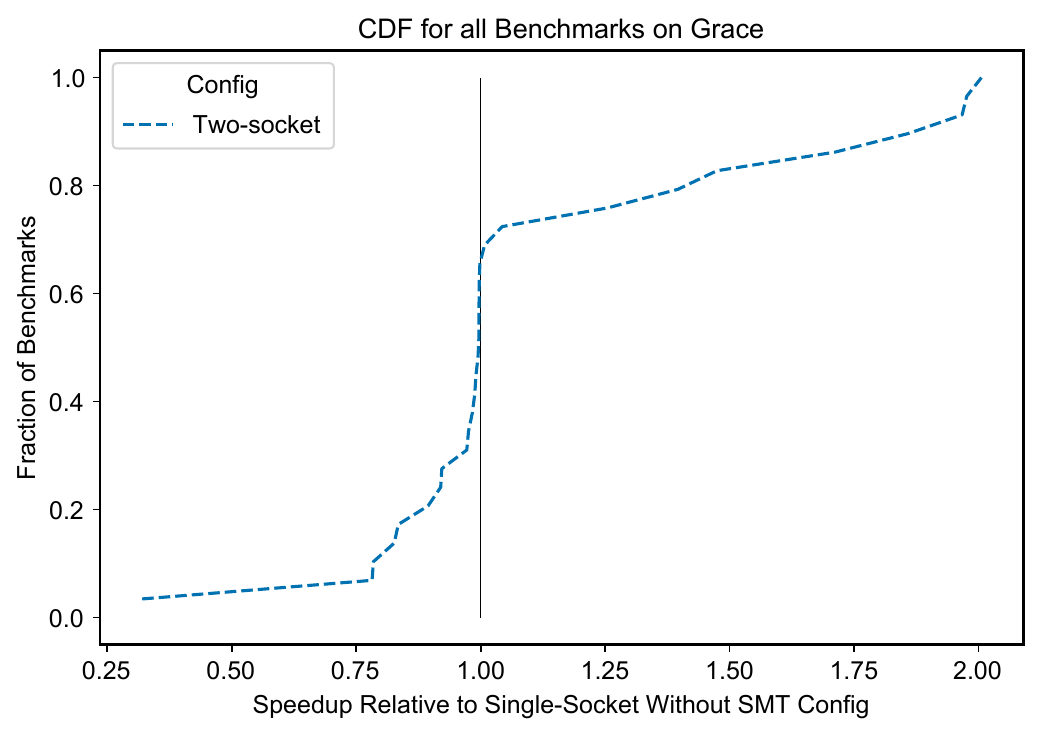}
  \end{subfigure}
  \caption{Cumulative Distribution Functions of Speedup Relative to Single-Socket Without SMT on Architectures With SMT and NUMA Available}
  \label{fig:smtnuma}
\end{figure*}

When comparing the performance data for the various RISC-V platforms with the larger and more expensive x86 and ARM machines, a clear question that arises is how the presence of simultaneous multithreading (SMT) and distinct NUMA domains affect the performance results.
While synchronization time between cores on a high-core-count machine is always slightly non-uniform, none of the RISC-V machines tested have multiple sockets, nor do any of them support SMT.
To better understand how both SMT and NUMA affect the results of these benchmarks on the machines used as baselines, we reran the suite of Chapel CLBG entries on the x86 and ARM machines with Chapel set to use:
\begin{itemize}
    \item A single socket without SMT enabled
    \item A single socket with SMT enabled
    \item All sockets without SMT enabled (the default)
    \item All sockets with SMT enabled
\end{itemize}

To effectively visualize the results, we plotted the cumulative distribution function of the speedup relative to a single socket without SMT enabled for each of the ARM and x86 machines.
These plots are shown in Figure~\ref{fig:smtnuma}.
These results highlight that SMT and NUMA effects can be helpful or harmful depending on the machine and benchmark.
This is unsurprising, given that some of the benchmarks tested are serial and others are synchronization-limited.
That said, this data also shows that the Zen 3 and Granite Rapids systems are more likely to scale gracefully to multiple chips than the Ice Lake and Grace systems.
We ultimately elected to use the whole-machine benchmarks as a baseline for evaluating the RISC-V machines since whole-machine performance is what matters most in the HPC space.


\section{Related Work}
\label{section:related}
RISC-V has attracted growing interest within the HPC community, particularly following the emergence of the first desktop-class systems.
Brown et al.~\cite{10.1145/3703001.3724388,brown2025} evaluated the NAS Parallel Benchmarks (NPB) and the RAJAPerf benchmark suite on the Milk-V Pioneer.
Diehl et al.~\cite{diehl2024preparing} assessed the performance of the astrophysics application Octo-Tiger using the Kokkos and HPX programming models.
AI workloads on RISC-V systems were investigated by Malenza et al.~\cite{malenza2025analysis} and Strack et al.~\cite{strack2026}.
In addition, Berger-Vergiat et al.~\cite{berger2023evaluation} and Barai et al.~\cite{barai2025bridging} used FireSim to emulate RISC-V hardware and evaluate representative HPC workloads.

Collectively, these studies focus on assessing the performance and maturity of HPC applications, programming models, and benchmark suites on existing or emulated RISC-V platforms.
A more comprehensive evaluation of RISC-V for production-scale HPC will require systems with HPC-class interconnects, such as InfiniBand or HPE Slingshot, as well as support for GPU accelerators and other heterogeneous computing technologies.

\section{Conclusion}
\label{section:conclusion}
Here we have discussed how Qthreads and Chapel were ported to RISC-V.
These experiences have highlighted the expanding need for performance portability in addition to traditional architecture portability.
We have also used a well-maintained portion of Chapel's benchmark suite to evaluate the current status of some available RISC-V hardware relative to existing x86 and ARM server-class systems.
While the RISC-V systems are not yet competitive with server-class hardware, they are also far less expensive.
These results provide hope that recent announcements regarding RISC-V hardware for server-class systems can result in competitive performance and widespread use in HPC.

\section*{Supplementary materials }
The benchmark scripts and performance data are available on GitHub\footnote{\url{https://github.com/diehlpkpapers/chapel-riscv}} or Zenodo~\footnote{\url{https://doi.org/10.5281/zenodo.21844540}}.

\section*{Acknowledgments}
Thanks to Dr. Amelia Henriksen for consulting on data visualization and analysis.

This work was supported by the U.S. Department of Energy through the Los Alamos National Laboratory. Los Alamos National Laboratory is operated by Triad National Security, LLC, for the National Nuclear Security Administration of the U.S. Department of Energy (Contract No. 89233218CNA000001). Approved by LANL as \textit{LA-UR-26-26953}. 

This article has been authored by an employee of National Technology~\& Engineering Solutions of Sandia, LLC under Contract No. DE-NA0003525 with the U.S. Department of Energy (DOE).
The employee owns all right, title and interest in and to the article and is solely responsible for its contents. The United States Government retains and the publisher, by accepting the article for publication, acknowledges that the United States Government retains a non-exclusive, paid-up, irrevocable, world-wide license to publish or reproduce the published form of this article or allow others to do so, for United States Government purposes. The DOE will provide public access to these results of federally sponsored research in accordance with the DOE Public Access Plan \url{https://www.energy.gov/downloads/doe-public-access-plan}.

\bibliographystyle{IEEEtran}
\bibliography{main}

\end{document}